%% file: E3.tex
\documentclass{aastex7}
\usepackage{ulem}

\newcommand\subs[1]{\textsubscript{#1}}
\newcommand\sups[1]{\textsuperscript{#1}}
\newcommand\rh[1]{\textcolor{black}{{\textit{r}\subs{H}}#1}}

\newcommand\trot[1]{\textcolor{black}{{\textit{T}\subs{rot}}#1}}

\newcommand\kms[1]{\textcolor{black}{{km\,s$^{-1}$}#1}}
\newcommand\ps[1]{\textcolor{black}{{s$^{-1}$}#1}}

\received{2026 February 6}
\revised{}
\accepted{}
\submitjournal{Planetary Science Journal}

\shorttitle{IRTF Studies of Comet C/2022 E3}
\shortauthors{Roth et al.}

\begin{document}

\title{Coma Volatile Composition and Thermal Physics in Comet C/2022 E3 (ZTF) Measured Near Closest Approach to Earth with NASA-IRTF}

\correspondingauthor{Nathan X. Roth}
\email{nathaniel.x.roth@nasa.gov}

\author[0000-0002-6006-9574]{Nathan X. Roth}
\altaffiliation{Visiting Astronomer at the Infrared Telescope Facility, which is operated by the University of Hawaii under contract NNH14CK55B with the National Aeronautics and Space Administration.}
\affiliation{Solar System Exploration Division, Astrochemistry Laboratory Code 691, NASA Goddard Space Flight Center, 8800 Greenbelt Rd., Greenbelt, MD 20771, USA}
\affiliation{Department of Physics, American University, 4400 Massachusetts Ave NW, Washington, DC 20016, USA}
\email{nathaniel.x.roth@nasa.gov}

\author[0000-0001-8843-7511]{Michael A. DiSanti}
\altaffiliation{Visiting Astronomer at the Infrared Telescope Facility, which is operated by the University of Hawaii under contract NNH14CK55B with the National Aeronautics and Space Administration.}
\affiliation{Solar System Exploration Division, Planetary Systems Laboratory Code 693, NASA-Goddard Space Flight Center, 8800 Greenbelt Rd., Greenbelt, MD 20771, USA}
\email{madman1954@comcast.net}

\author[0000-0002-6391-4817]{Boncho P. Bonev}
\altaffiliation{Visiting Astronomer at the Infrared Telescope Facility, which is operated by the University of Hawaii under contract NNH14CK55B with the National Aeronautics and Space Administration.}
\affiliation{Department of Physics, American University, 4400 Massachusetts Ave NW, Washington, DC 20016, USA}
\email{bonev@american.edu}

\author{Neil Dello Russo}
\altaffiliation{Visiting Astronomer at the Infrared Telescope Facility, which is operated by the University of Hawaii under contract NNH14CK55B with the National Aeronautics and Space Administration.}
\affiliation{Johns Hopkins University Applied Physics Laboratory, 11100 Johns Hopkins Rd., Laurel, MD, 20723 USA}
\email{neil.dello.russo@jhuapl.edu}

\author[0000-0003-0142-5265]{Erika L. Gibb}
\altaffiliation{Visiting Astronomer at the Infrared Telescope Facility, which is operated by the University of Hawaii under contract NNH14CK55B with the National Aeronautics and Space Administration.}
\affiliation{Department of Mathematics, Physics, Astronomy, and Statistics, 1 University Blvd., University of Missouri-St.Louis, St. Louis, MO, 63121 USA}
\email{gibbe@umsl.edu}

\author[0000-0002-8227-9564]{Ronald J. Vervack, Jr.}
\altaffiliation{Visiting Astronomer at the Infrared Telescope Facility, which is operated by the University of Hawaii under contract NNH14CK55B with the National Aeronautics and Space Administration.}
\affiliation{Johns Hopkins University Applied Physics Laboratory, 11100 Johns Hopkins Rd., Laurel, MD, 20723 USA}
\email{ron.vervack@jhuapl.edu}

\author[0000-0003-2277-6232]{Mohammad Saki}
\altaffiliation{Visiting Astronomer at the Infrared Telescope Facility, which is operated by the University of Hawaii under contract NNH14CK55B with the National Aeronautics and Space Administration.}
\affiliation{Department of Mathematics, Physics, Astronomy, and Statistics, 1 University Blvd., University of Missouri-St.Louis, St. Louis, MO, 63121 USA}
\email{sakim@umsl.edu}

\author[0000-0002-0622-2400]{Adam J. McKay}
\affiliation{Department of Physics \& Astronomy, Appalachian State University, 525 Rivers Street, Boone, NC 28608  USA}
\email{mckayaj@appstate.edu}

\author[0000-0003-2011-9159]{Hideyo Kawakita}
\affiliation{Koyama Astronomical Observatory, Kyoto Sangyo University, Motoyama, Kamigamo, Kita-ku, Kyoto, 603-8555, Japan}
\email{hideyokawakita7504@gmail.com}

\author[0000-0001-7694-4129]{Stefanie N. Milam}
\affiliation{Solar System Exploration Division, Astrochemistry Laboratory Code 691, NASA Goddard Space Flight Center, 8800 Greenbelt Rd, Greenbelt, MD 20771, USA}
\email{stefanie.n.milam@nasa.gov}

\author{Martin A. Cordiner}
\affiliation{Solar System Exploration Division, Astrochemistry Laboratory Code 691, NASA Goddard Space Flight Center, 8800 Greenbelt Rd, Greenbelt, MD 20771, USA}
\affiliation{Department of Physics, Catholic University of America, Washington DC, USA}
\email{martin.cordiner@nasa.gov}

\author{K. D. Foster}
\affiliation{Department of Chemistry, University of Virginia, Charlottesville, VA 22904, USA}
\email{kfoster@virginia.edu}


\input{abstract}



\input{Introduction}

\input{Observations}

\input{Results}

\input{Discussion}

\input{Conclusions}

\begin{acknowledgments}
Data for this study were obtained at the NASA Infrared Telescope Facility (IRTF), operated by the University of Hawaii under contract NNH14CK55B with the National Aeronautics and Space Administration, as well as from the W. M. Keck Observatory, which is operated as a scientific partnership among the California Institute of Technology, the University of California, and the National Aeronautics and Space Administration. We are most fortunate to the have the opportunity to conduct observations from Maunakea, and recognize the very significant cultural role and reverence that the summit of Maunakea has always had within the indigenous community. NXR and MAD acknowledge support by the NASA Solar System Observations Program through 22-SSO22-0013, and NDR and RJV through 80NSSC22K1401. ELG and BPB acknowledge support by NSF grants AST-2009910, AST-2511626, AST-2009398, and AST-2511627 . BPB acknowledges support by NASA Solar System Workings (80NSSC20K0651).
\end{acknowledgments}

\software{Astropy \citep{astropy:2013, astropy:2018, astropy:2022},
Small-Bodies-Node/ice-sublimation \citep{VanSelous2021},
lmfit \citep{Newville2016}} 

\bibliography{E3}{}
\bibliographystyle{aasjournal}



\end{document}

%% file: abstract.tex
\begin{abstract}

The 2023 perihelion passage of comet C/2022 E3 (ZTF) afforded an opportunity to measure the abundances and spatial distributions of coma volatiles in an Oort cloud comet at high spatial resolution near its close approach to Earth ($\Delta_\mathrm{min}\sim 0.28$ au on UT February 1). We conducted near-infrared 
spectroscopic observations of C/2022 E3 (ZTF) using iSHELL at the NASA Infrared Telescope Facility on UT 2023 February 9. Our measurements securely detected fluorescent emission from H$_2$O, CO, OCS, CH$_3$OH, CH$_4$, C$_2$H$_6$, C$_2$H$_2$, and HCN. For each instrumental setting we took exposures with the slit oriented parallel and also perpendicular to the projected Sun-comet vector, thereby enabling a test of the spatial distributions of these molecules. We report rotational temperatures, production rates, and abundance ratios (i.e., mixing ratios) for all sampled species. Our measurements found that molecular abundances in C/2022 E3 were depleted compared to their average values in Oort cloud comets with the exception of OCS, which was consistent. The H$_2$O production rate varied significantly and was likely tied to nucleus rotation effects. Measurements at the two slit orientations showed distinct column density and rotational temperature profiles for H$_2$O. Peak temperatures occurred off-nucleus and slower cooling was present in the anti-sunward hemisphere, consistent with the presence of icy grain sublimation in the coma. 

\end{abstract}

%% file: Introduction.tex
\section{Introduction} \label{sec:intro}

Remote sensing of cometary atmospheres provides a window into the chemistry and physics that were preserved in their nuclei from the nascent solar system \citep{Bockelee2004,DelloRusso2016a}. In particular, measurements with state-of-the-art long-slit infrared spectrographs, such as iSHELL at the NASA Infrared Telescope Facility (IRTF), enable spatial-spectral studies of coma chemistry. Molecular production rates (and relative abundances) and rotational temperatures for a suite of species can be retrieved by sampling multiple ro-vibrational transitions simultaneously with iSHELL's cross-dispersed capabilities \citep{DiSanti2017,Faggi2019}. Characterizing the trends in emission intensity along the 15$\arcsec$ long slit reveals the projected spatial distribution of each molecule in the coma and affords the opportunity to test coma thermal physics by mapping variations in rotational temperature with nucleocentric distance \citep{Bonev2013,Bonev2014}. 

The close approach of comet C/2022 E3 (ZTF; hereafter E3) in 2023 February was an excellent case for such a study. Owing to its intrinsic brightness and anticipated close approach to Earth, E3 was studied from the radio to the optical, including with ALMA and JWST \citep{Foster2026}. Its minimum geocentric distance ($\Delta$) of 0.28 au on 2023 February 1 enabled studies of the distributions of molecular abundances and temperatures in its inner coma ($\sim750$ km from the nucleus). Here we report observations of E3 on 2023 February 9 taken at a heliocentric distance (\rh{}) of 1.12 au and $\Delta=0.38$ au. We detected emission from H$_2$O, CO, OCS, CH$_3$OH, CH$_4$, C$_2$H$_6$, C$_2$H$_2$, HCN, and OH$^*$ (prompt emission). The high geocentric velocity of the comet ($d\Delta/dt\sim38$ \kms{}) enabled measures of the hypervolatile species CO and CH$_4$ by Doppler shifting their emissions away from (highly opaque) telluric cores to regions of favorable atmospheric transmittance. We performed compositional studies of E3 using three iSHELL instrument settings. We performed multiple integrations for each setting, alternating the orientation of the slit between position angles parallel and perpendicular to the projected Sun-comet line. We provide molecular production rates, abundances, and rotational temperatures for each detected species. We provide spatial profiles of molecular emission along the slit for H$_2$O, HCN, CO, C$_2$H$_6$, and CH$_4$, and of variations in rotational temperature along the slit for H$_2$O. In Section~\ref{sec:obs} we detail our observations and data reduction. In Section~\ref{sec:results} we present our results. In Section~\ref{sec:discussion} we discuss the implications of our results for comet E3 and place them into context with the larger comet population.

%% file: Observations.tex
\section{Observations and Data Reduction} \label{sec:obs}
E3 is an Oort cloud comet (OCC) that reached perihelion (\textit{q} = 1.11 au) on UT 2023 January 12. As part of a larger observing campaign throughout 2023 January and February, we targeted E3 on 2023 February 9 with the high-resolution, near-infrared facility spectograph iSHELL \citep{Rayner2022} at the 3 m NASA-IRTF to characterize its volatile composition. The observing log is shown in Table~\ref{tab:obslog}. We utilized the Lp1 and M2 iSHELL settings, along with a custom L-band setting spanning 2.81-3.09 $\mu$m (see Table~\ref{tab:obslog}), so as to fully sample a suite of molecular abundances. We alternated the orientation of the slit between successive observations of a given instrumental setting. 

\begin{deluxetable*}{cccccccccc}
\tablenum{1}
\tablecaption{Observing Log\label{tab:obslog}}
\tablewidth{0pt}
\tablehead{
\colhead{UT Time} & \colhead{Setting} &\colhead{Target} & \colhead{\textit{T}\subs{int}} & \colhead{\textit{r}\subs{H}} & \colhead{$\Delta$} & \colhead{d$\Delta$/dt} & \colhead{Molecules} & \colhead{Slit PA} \\
\colhead{} & \colhead{} & \colhead{} & \colhead{(min)}  & \colhead{(au)} & \colhead{(au)} & \colhead{(km s\sups{-1})} & \colhead{Sampled} & \colhead{($\degr$)}
}
\startdata
\multicolumn{9}{c}{UT 2023 February 9} \\
\hline
03:07 -- 03:52 & L-Custom & E3 & 40 & 1.197 & 0.375 & 37.4 & H\subs{2}O, HCN, C$_2$H$_2$, OH$^*$ & 88 \\
04:10 -- 04:44 &  &  & 40 & 1.197 & 0.376 & 37.6 &  & 178 \\
05:08 -- 05:21 &  & BS-1040 & -- & -- & -- & -- & -- & --\\
05:22 -- 05:37 & Lp1 & BS-1040 & -- & -- & -- & -- & -- & --\\
05:38 -- 05:50 & M2 & BS-1040 & -- & -- & -- & -- & -- & --\\
05:54 -- 06:41 &  & E3 & 40 & 1.198 & 0.378 & 38.0 & H$_2$O, CO, OCS, CN & 88 \\
06:54 -- 07:41 & & E3 & 40 & 1.198 & 0.379 & 38.2 & & 178 \\
07:55 -- 08:40 & Lp1 & E3 & 40 & 1.198 & 0.379 & 38.4 & C$_2$H$_6$, CH$_3$OH, CH$_4$, OH$^*$ & 178 \\
08:47 -- 09:32 & & & 40 & 1.198 & 0.380 & 38.6 & & 88 \\
09:39 -- 10:15 & & & 32 & 1.199 & 0.381 & 38.7 & & 178 \\
10:25 -- 10:34 & & BS-2560 & -- & -- & -- & -- & -- & --\\
10:41 -- 10:49 & M2 & BS-2560 & -- & -- & -- & -- & -- & --\\ 
10:56 -- 11:02 & L-Custom & BS-2560 & -- & -- & -- & -- & -- & --\\
\enddata
\tablecomments{\textit{r}\subs{H}, $\Delta$, and d$\Delta$/dt are the heliocentric distance, geocentric distance, and geocentric velocity, respectively, of E3 at the time of observations. \textit{T}\subs{int} is the integrated time on-source. L-Custom is a custom iSHELL setting spanning 2.81 -- 3.09 $\mu$m. The solar phase angle ($\phi$) was $48\degr$. Seeing was $\sim0\farcs7$. 
}
\end{deluxetable*}

Observations were performed with a 6 pixel (0$\farcs$75) wide slit with resolving power ($\lambda/\Delta\lambda) \sim 4.5\times10^4$. We used a standard ABBA nod pattern in which the telescope is nodded along the slit between successive exposures, thereby placing the comet at two distinct positions along the slit (``A'' and ``B'') in order to facilitate sky subtraction. The A and B beams were symmetrically placed about the midpoint along the 15$\arcsec$ long slit and separated by half its length. E3 was bright and easily acquired with iSHELL's near-infrared active guiding system. Combining spectra of the nodded beams as A-B-B$+$A canceled emissions from thermal background, instrumental biases, and sky emission (lines and continuum) to second order in air mass. Flux calibration was performed using an appropriately placed bright infrared flux standard star using a wide (4$\farcs$0) slit. 

We employed data reduction procedures that have been rigorously tested and are described extensively in the refereed literature \citep{Bonev2005,DiSanti2006,DiSanti2014,Villanueva2009,Radeva2010}, including their application to unique aspects of iSHELL spectra \citep{DiSanti2017,Roth2020}. Each echelle order within an iSHELL setting was processed individually as previously described, such that each row corresponded to a unique position along the slit, and each column to a unique wavelength. Spectra were extracted from the processed frames by summing the signal over 15 rows (approximately 2$\farcs$5), seven rows to each side of the nucleus, defined as the peak of dust emission in a given spectral order. 

We determined contributions from continuum and gaseous emissions in comet spectra as previously described \citep[e.g.,][]{DiSanti2016,DiSanti2017} and illustrate the procedure in Figure~\ref{fig:h2o}. We retrieved burdens of telluric absorbers using the NASA Planetary Spectrum Generator \citep[PSG;][]{Villanueva2018}. We convolved the fully resolved transmittance function to the resolving power of the data ($\sim 4.5 \times 10^4$) and scaled it to the level of the comet continuum. We then subtracted the modeled continuum to isolate cometary emission lines and compared synthetic models of fluorescent emission for each targeted species to the observed line intensities.

Nucleocentric (or nucleus-centered) production rates (\textit{Q}\subs{NC}) were determined using a well-documented formalism \citep{DelloRusso1998,DiSanti2001,Bonev2005,Villanueva2011a}; see Section 3.2.2 of \cite{DiSanti2016} for further details. The \textit{Q}\subs{NC} were multiplied by an appropriate growth factor (GF), determined using the \textit{Q}-curve methodology \citep[e.g.,][]{DelloRusso1998,DiSanti2001,Bonev2005,Gibb2012} to establish total (or global) production rates (\textit{Q}). The $Q$-curve formalism
corrects for atmospheric seeing, which suppresses signal along lines of sight passing close to the nucleus owing to the use of a narrow slit, as well as for potential perpendicular drift of the comet during an exposure sequence. 

We observed appropriately placed bright infrared flux standards near the beginning and at the end of the observing run so as to provide robust flux calibration and to test for changing conditions. NASA PSG fits to telluric H$_2$O absorption lines provided a measure of the precipitable water vapor during our flux standard observations: 0.92 mm for BS-1040 (UT 05:08--05:50) and 1.92 mm for BS-2560 (UT 10:25--11:02). This indicates that the telluric H$_2$O burden increased over the course of our observations. However, flux calibration factors ($\Gamma$, W/m$^2$/cm$^{-1}$/(counts/s)) derived from measurements of BS-2560 at the end of the observations were only 4\% lower than those derived from BS-1040 near the beginning of the observations. Given this small difference  and to facilitate a uniform comparison, we applied flux calibration factors from BS-1040 to the entire data set. We also note that any change in applied flux calibration factors would only affect absolute $Q$'s, not molecular abundances for molecules within the same setting.

We estimate the uncertainty in \textit{Q} due to flux calibration within an instrumental setting to be 3\% for the L-Custom setting, 2\% for the Lp1 setting, and 3\% for the M2 setting based on the standard deviation of flux calibration factors ($\Gamma$) within each echelle order taken from eight exposures. We incorporated this additional uncertainty into our production rates. Global production rates for all detected molecules are listed in Table~\ref{tab:comp}. GFs were determined for H$_2$O, HCN, CO, CH$_4$, C$_2$H$_6$, and CH$_3$OH. 

\subsection{Mixing Ratios of Volatile Species}\label{subsec:mixing}
\subsubsection{Molecular Fluorescence Analysis}\label{subsubsec:fluorescence}
Synthetic models of fluorescent emission for each targeted species were compared to observed line intensities, after correcting each modeled \textit{g}-factor (line intensity) for the monochromatic atmospheric transmittance at its Doppler-shifted wavelength (according to the geocentric velocity of the comet at the time of the observations). The \textit{g}-factors used in synthetic fluorescent emission models in this study were generated with quantum mechanical models from the PSG. A Levenburg-Marquardt nonlinear minimization technique \citep{Villanueva2008} was used to fit fluorescent emission from all species simultaneously in each echelle order, allowing for high precision results, even in spectrally crowded regions containing many spectral lines within a single instrumental resolution element. Production rates for each sampled species were determined from the appropriate fluorescence model at the rotational temperature of each molecule (Section~\ref{subsubsec:trot}).

\subsubsection{Determination of Rotational Temperature}\label{subsubsec:trot}
Rotational temperatures (\textit{T}\subs{rot}) were determined using correlation and excitation analyses as described in \cite{Bonev2005,Bonev2008,DiSanti2006,Villanueva2008}. In general, well-constrained rotational temperatures can be determined for individual species having intrinsically bright lines and for which a sufficiently broad range of excitation energies is sampled. Utilizing the large spectral grasp of iSHELL, these conditions were satisfied in nucleus-centered spectra for H$_2$O, CO, OCS, CH$_4$, C$_2$H$_6$, and CH$_3$OH, and for off-nucleus spectra of H$_2$O. 

Nucleus-centered rotational temperatures for a given molecule were in formal agreement within an instrumental setting. For instance, \trot{} for H$_2$O in the L-custom integration measured parallel to the Sun-comet line agreed with that measured perpendicular to the Sun-comet line within $1\sigma$, and the same was true for HCN. However, \trot{} for H$_2$O showed modest variation between settings, rising from $78\pm1$ K in both L-custom observations to $87\pm5$ K and $90\pm5$ K in the M2 integrations. 

In terms of the trace species, \trot{} was nominally lower for CO ($77\pm10$ K and $69\pm7$ K measured perpendicular and parallel to the Sun-comet line, respectively) and OCS ($77\pm21$ K) than for H$_2$O measured in the M2 observations, and \trot{} for HCN ($108\pm13$ K, $116\pm14$ K measured parallel and perpendicular to the Sun-Comet line, respectively) was higher than H$_2$O in the L-Custom setting. \trot{} for CH$_4$ and C$_2$H$_6$ were in agreement for all three Lp1 observations and with that for H$_2$O measured in L-custom and M2, whereas CH$_3$OH was nominally colder. However, all rotational temperatures agree with one another within $2\sigma$. This is consistent with previous work demonstrating that rotational temperatures for primary species sampled at near-infrared wavelengths are generally in agreement \citep[see for example][and references therein]{Gibb2012,DiSanti2016}. All rotational temperatures are tabulated in Table~\ref{tab:comp}.

\begin{deluxetable*}{ccccccc}
\tablenum{2}
\tablecaption{Molecular Composition of C/2022 E3 as Measured by iSHELL\label{tab:comp}}
\tablewidth{0pt}
\tablehead{
\colhead{Setting} & \colhead{Species} & \colhead{\textit{T}\subs{rot}\sups{a}} & \colhead{GF\sups{b}} & \colhead{\textit{Q}\sups{c}} & \multicolumn{2}{c}{Relative Abundance} \\
\colhead{} & \colhead{ } & \colhead{(K)} & \colhead{ } & \colhead{(10$^{25}$ s\sups{-1})}  & \colhead{\textit{Q}\subs{x}/\textit{Q}\subs{H2O}\sups{d} (\%)} &  \colhead{\textit{Q}\subs{x}/\textit{Q}\subs{C2H6}\sups{e}} 
}
\startdata
\multicolumn{7}{c}{UT 2023 February 9, \textit{r}\subs{H} = 1.197 au, $\Delta$ = 0.375 au} \\
\hline
L-Custom       & H$_2$O & $78\pm1$ & $2.21\pm 0.06$ & $8352\pm150$ & \textbf{100} & $374\pm11$ \\
PA = 88$\degr$ & HCN & $108\pm13$ & $2.05\pm0.22$ & $8.12\pm0.53$ & \textbf{$0.097\pm0.007$} & $0.36\pm0.03$ \\
 & C$_2$H$_2$ & (78) & (2.21) & $4.45\pm0.86$ & \textbf{$0.05\pm0.01$} & $0.20\pm0.04$ \\
\hline
L-Custom       & H$_2$O & $78\pm1$ & $2.38\pm 0.02$ & $8050\pm102$ & \textbf{100} & $361\pm9$ \\
PA = 178$\degr$ & HCN & $116\pm14$ & $1.73\pm0.21$ & $6.26\pm0.53$ & \textbf{$0.078\pm0.007$} & $0.28\pm0.03$ \\
 & C$_2$H$_2$ & (78) & (2.38) & $<3.45 (3\sigma)$ & \textbf{$<0.04 (3\sigma)$} & $<0.16 (3\sigma$) \\
\hline
M2 & H$_2$O & $87\pm5$ & $2.03\pm0.15$ & $6080\pm228$ & \textbf{100} & $273\pm13$ \\
PA = 178$\degr$ & CO & $77\pm10$ & $2.06\pm0.15$ & $33.3\pm2.6$ & \textbf{$0.55\pm0.05$} & $1.49\pm0.13$ \\
\hline
M2 & H$_2$O & $90\pm5$ & $2.37\pm0.11$ & $6629\pm259$ & \textbf{100} & $297\pm14$ \\
PA = 88$\degr$ & CO & $69\pm7$ & $2.42\pm0.14$ & $35.9\pm2.3$ & \textbf{$0.54\pm0.04$} & $1.61\pm0.12$ \\
\hline
M2$^{(a)}$ & OCS & $77\pm21$ & (2.03) & $3.06\pm0.53$ & \textbf{$0.046\pm0.008$} & $0.14\pm0.03$ \\
\hline
Lp1 & OH$^*$ & (78) & $2.01\pm0.12$ & $7159\pm253$ & 100 & \textbf{$262\pm13$} \\
PA=178$\degr$ & CH$_4$ & $70\pm10$ & $2.27\pm0.12$ & $18.6\pm1.3$ & $0.26\pm0.02$ & \textbf{$0.68\pm0.06$} \\
    & C$_2$H$_6$ & $77\pm7$ & $2.54\pm0.05$ & $27.2\pm0.9$ & $0.38\pm0.01$ & \textbf{1} \\
    & CH$_3$OH & $88\pm17$ & (2.01) & $58.5\pm6.3$ & $0.82\pm0.04$ & \textbf{$2.15\pm0.24$} \\
\hline
Lp1 & OH$^*$ & (78) & $2.77\pm0.16$ & $8966\pm883$ & 100 & \textbf{$416\pm43$} \\
PA=88$\degr$ & CH$_4$ & $87\pm12$ & $2.50\pm0.28$ & $25.9\pm1.0$ & $0.29\pm0.03$ & \textbf{$1.20\pm0.06$} \\
    & C$_2$H$_6$ & $70\pm7$ & $2.56\pm0.08$ & $21.5\pm0.7$ & $0.24\pm0.03$ & \textbf{1} \\
    & CH$_3$OH & $64\pm9$ & (2.77) & $53.7\pm4.5$ & $0.60\pm0.08$ & \textbf{$2.50\pm0.22$} \\
\hline
Lp1 & OH$^*$ & (78) & $2.32\pm0.15$ & $7064\pm352$ & 100 & \textbf{$339\pm20$} \\
PA=178$\degr$ & CH$_4$ & $76\pm13$ & $2.58\pm0.51$ & $19.4\pm1.5$ & $0.28\pm0.03$ & \textbf{$0.93\pm0.08$} \\
    & C$_2$H$_6$ & $70\pm7$ & $2.32\pm0.08$ & $20.8\pm0.6$ & $0.30\pm0.02$ & \textbf{1} \\
    & CH$_3$OH & $49\pm9$ & (2.32) & $45.4\pm4.8$ & $0.64\pm0.08$ & \textbf{$2.18\pm0.24$} \\
\enddata
\tablecomments{\sups{a}Rotational temperature. \sups{b}Growth factor. This represents the ratio of global \textit{Q} to \textit{Q}\subs{NC} based on the central $0\farcs75\times0\farcs83$ aperture having peak emission flux. \sups{c}Production rate. An additional uncertainty of 3\%, 2\%, and 3\% was incorporated into \textit{Q}'s for L-Custom, M2, and Lp1, respectively, based on the standard deviation of flux calibration factors ($\Gamma$, W/m$^2$/cm$^{-1}$/(count/s)). \sups{d}Mixing ratio with respect to simultaneously measured H\subs{2}O (H$_2$O = 100).
\sups{e}Mixing ratio with respect to C\subs{2}H\subs{6} (C$_2$H$_6$ = 1). A weighted average $Q$(C$_2$H$_6$)=$(2.23\pm0.04)\times10^{26}$ \ps{} calcaulated from all three Lp1 observations was used to calculate mixing ratios for the Lcustom and M2 settings, and an additional 10\% uncertainty was incorporated to account for the non-simultaneous observations of C$_2$H$_6$. In the context of interpreting relative abundances, bold font emphasizes that H$_2$O was only directly measured in the L-Custom and M2 settings, whereas C$_2$H$_6$ was only directly measured in the Lp1 setting.
}
\end{deluxetable*}

%% file: Results.tex
\section{Results}\label{sec:results}
\begin{figure*}
\plotone{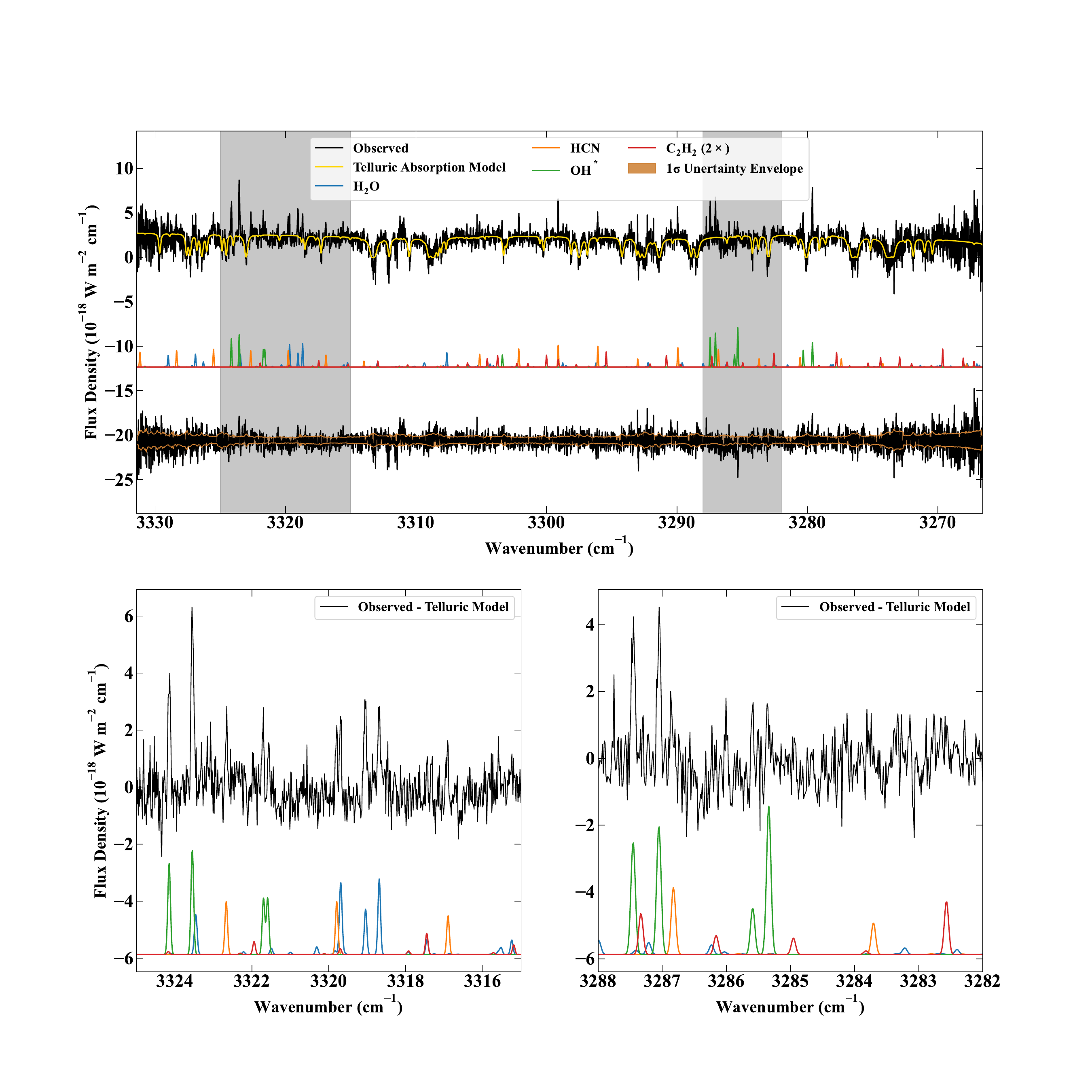}
\caption{\textbf{Upper.} Detections of H$_2$O, HCN, C$_2$H$_2$, and OH$^*$ in E3 covering iSHELL echelle order 170--172. The uppermost trace is the observed cometary spectrum with the telluric transmittance model superimposed. Below are individual molecular fluorescence models color-coded by species. The bottom trace shows the residual (cometary spectrum minus all models) with the $1\sigma$ uncertainty envelope overlaid and shaded. \textbf{Lower.} Zoomed plots covering the gray shaded regions indicated in the upper panel. In this case, the uppermost trace is the telluric-subtracted comet spectrum, with individual fluorescence models plotted below. We searched for NH$_2$, a potentially confounding species in this spectral region, but did not detect it. Our $3\sigma$ upper limit is NH$_2$/H$_2$O $<$ 0.03\%.
\label{fig:h2o}}
\end{figure*}

\begin{figure*}
\plotone{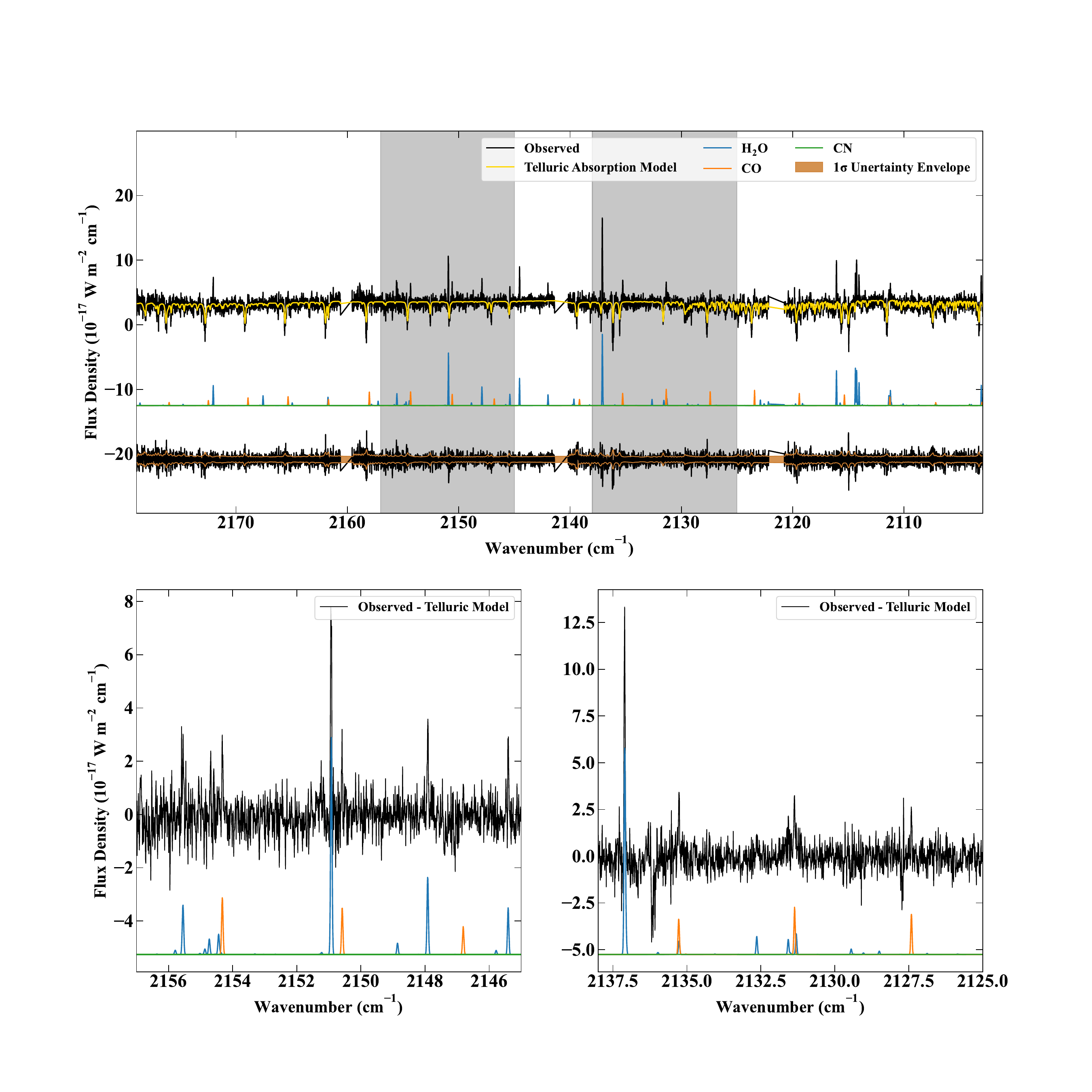}
\caption{\textbf{Upper, Lower.} Detections of H$_2$O and CO in E3 covering iSHELL echelle orders 109--112, with traces and labels as in Figure~\ref{fig:h2o}. 
\label{fig:co}}
\end{figure*}

\begin{figure*}
\plotone{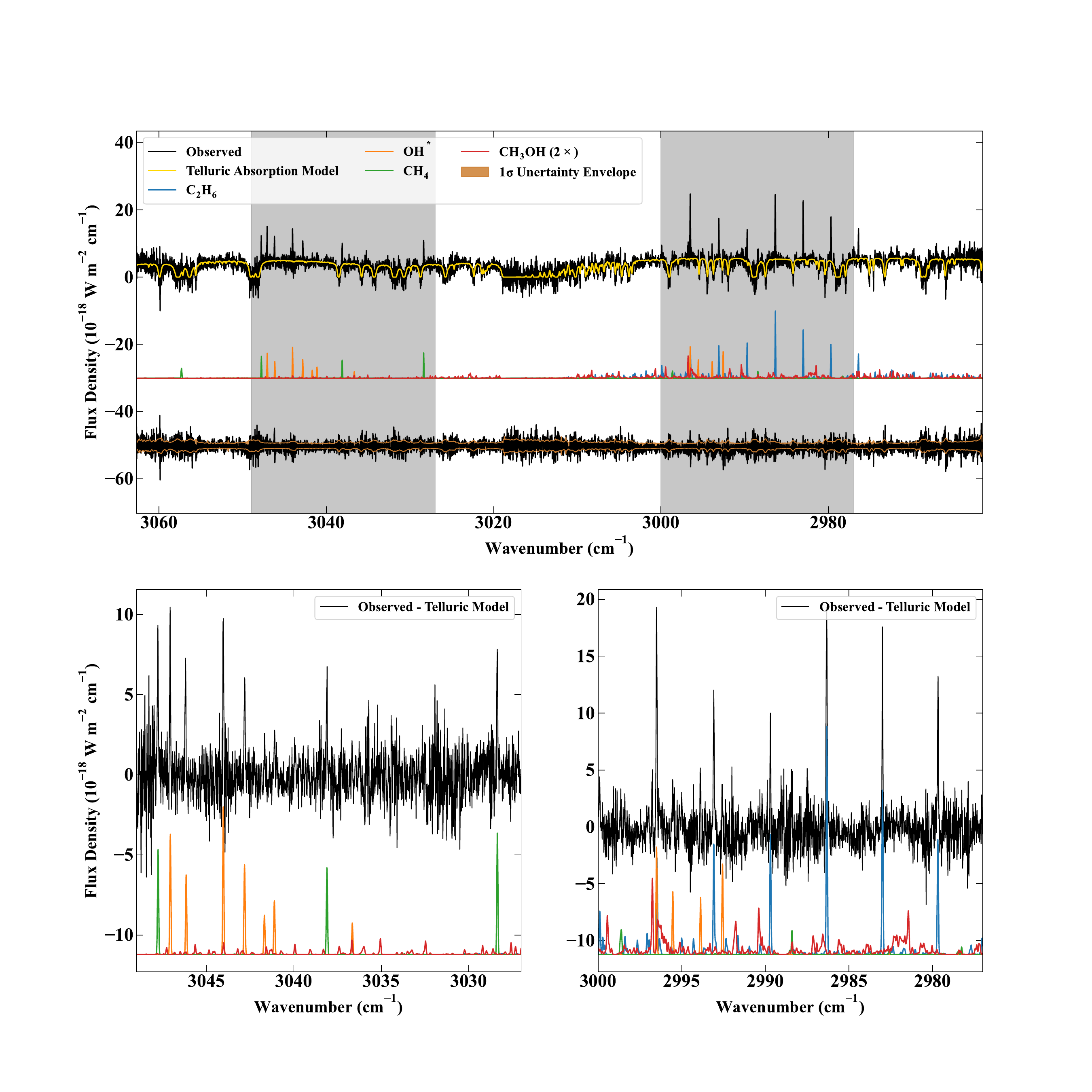}
\caption{\textbf{Upper, Lower.} Detections of H$_2$O (OH$^*$), C$_2$H$_6$, CH$_3$OH, and CH$_4$ in E3 covering iSHELL echelle orders 154--158, with traces and labels as in Figure~\ref{fig:h2o}.  
\label{fig:c2h6}}
\end{figure*}

\begin{figure*}
\plotone{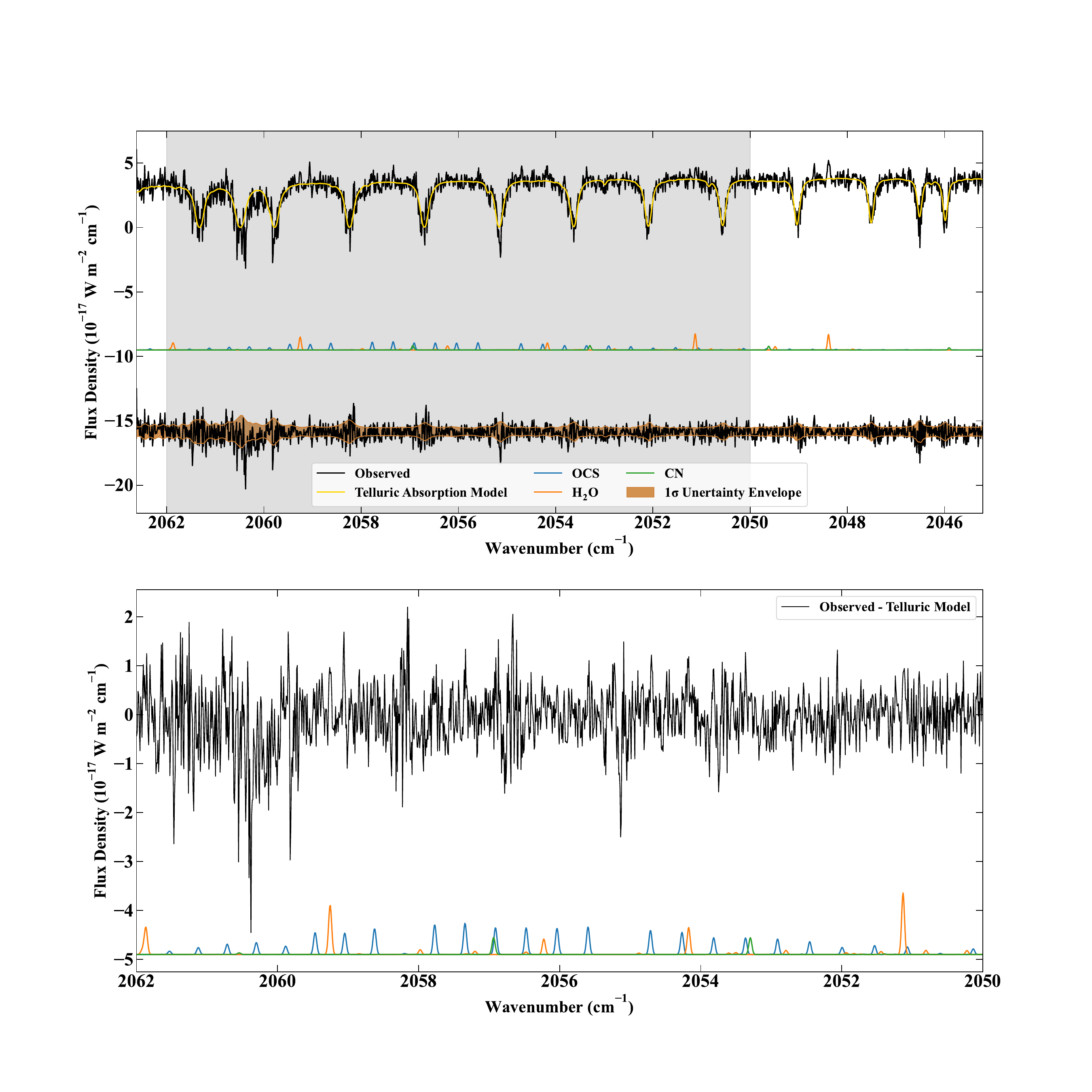}
\caption{\textbf{Upper, Lower.} Detections of H$_2$O, OCS, and CN in E3 covering iSHELL echelle order 106, with traces and labels as in Figure~\ref{fig:h2o}.  
\label{fig:ocs}}
\end{figure*}

We securely detected molecular emission from H$_2$O, CO, CH$_4$, C$_2$H$_6$, CH$_3$OH, HCN, C$_2$H$_2$, and OH* (prompt emission) in individual 40 minute integrations in the coma of E3. Figures~\ref{fig:h2o}, \ref{fig:co}, and \ref{fig:c2h6} show clear detections of multiple transitions of each detected species superimposed on the cometary continuum. By combining signal from both M2 integrations (80 minutes) simultaneously, we were also able to detect OCS (Figure~\ref{fig:ocs}).

\input{results-spatial}

\input{results-temperature}

%% file: results-spatial.tex
\subsection{Spatial Profiles of Emission}\label{subsec:profiles}
\begin{figure*}
\plotone{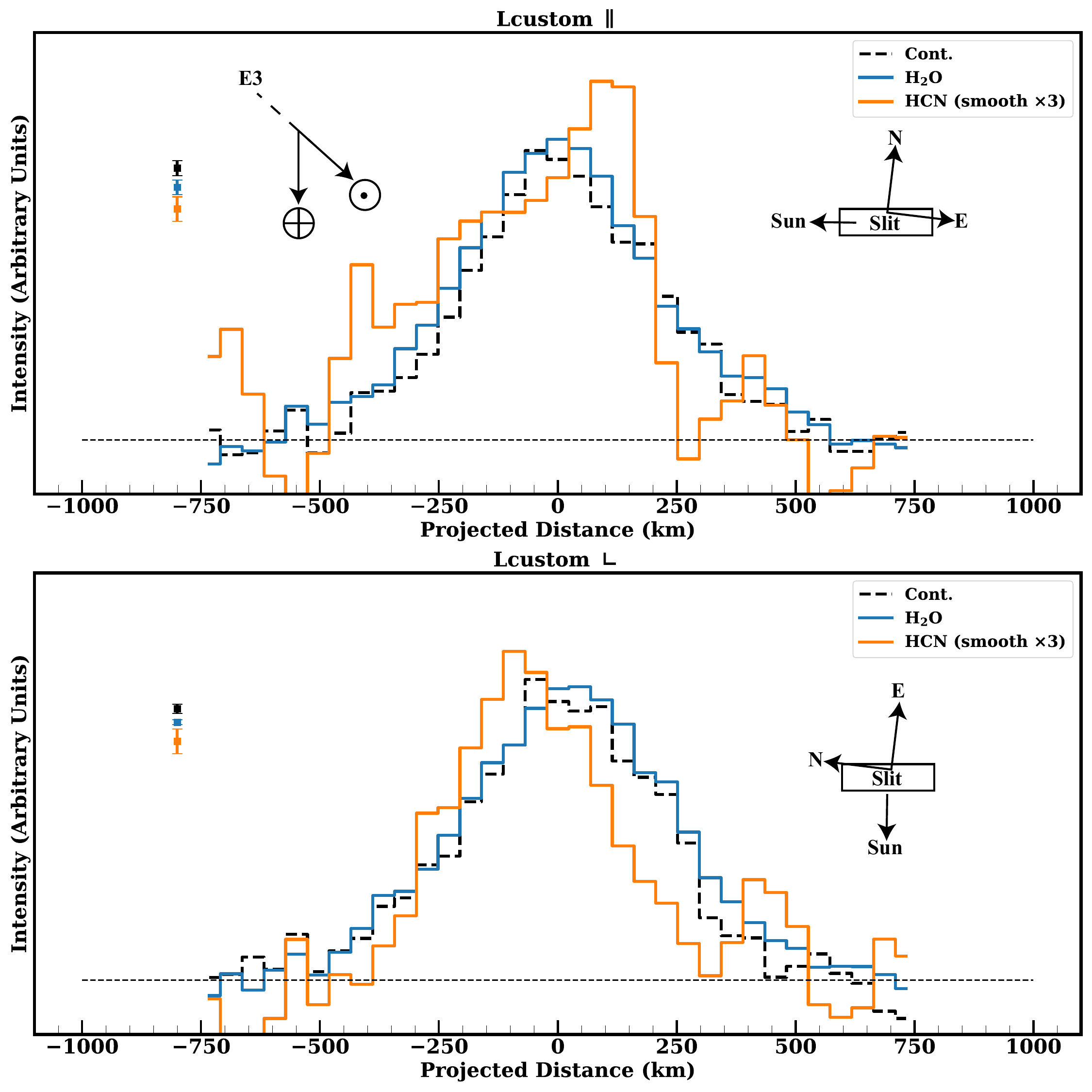}
\caption{\textbf{Upper.} Spatial profiles of emissions for co-measured H$_2$O (blue), HCN (orange), and continuum (dashed) in E3. Color-coded 1$\sigma$ uncertainties are shown for the profile of each species. The slit was oriented along the projected Sun-comet line (position angle 88$\degr$), with the Sun-facing direction to the left as indicated. Also shown is the Sun-comet-Earth angle (phase angle, $\beta$) of 48$\degr$. ``Smooth'' indicates the width (pixels) of a boxcar smoothing kernel applied to HCN. \textbf{Lower.} Spatial profiles of co-measured H$_2$O, HCN, and continuum in E3 with the slit oriented perpendicular to the projected Sun-comet line (position angle 178$\degr$).  
\label{fig:profiles1}}
\end{figure*}

\begin{figure*}
\plotone{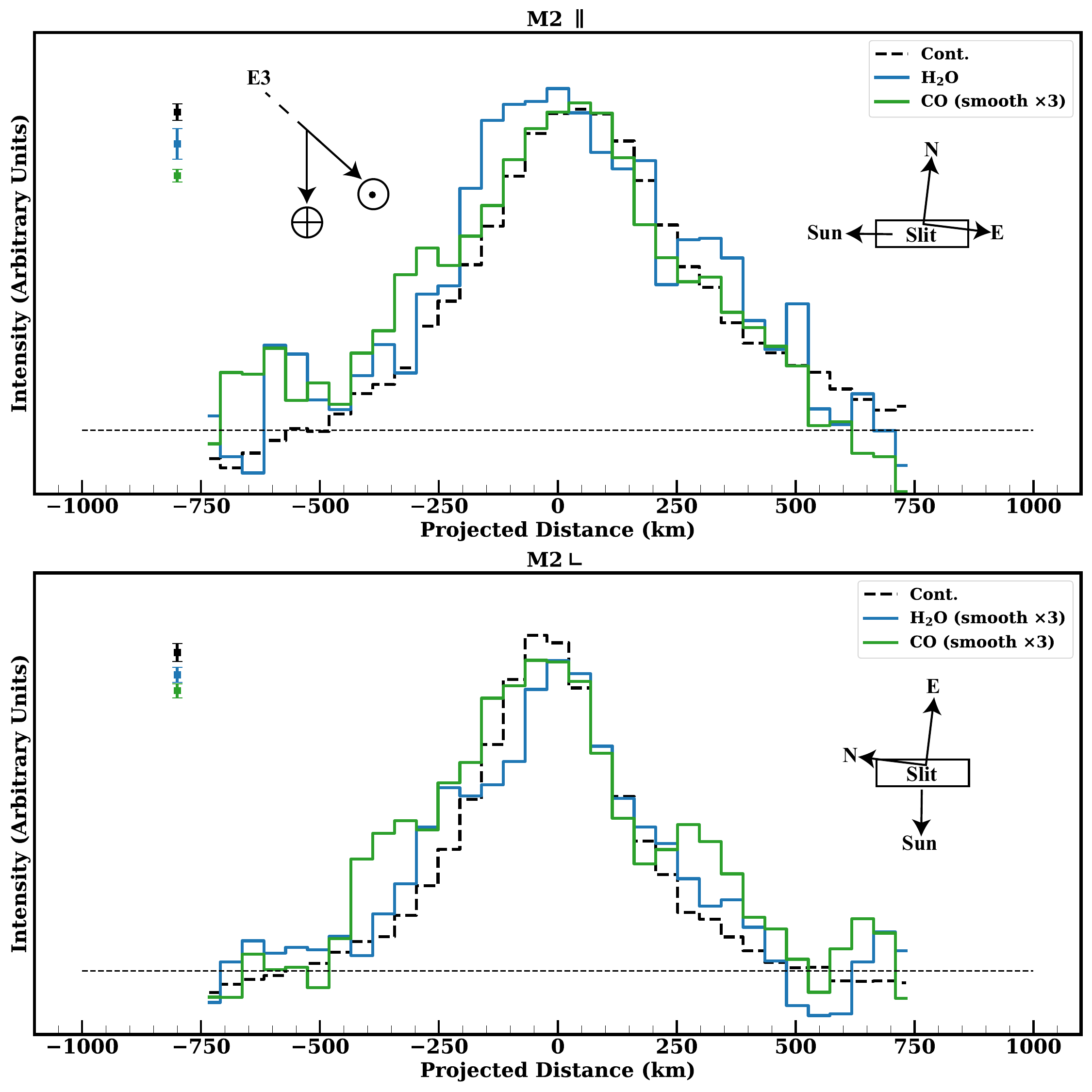}
\caption{\textbf{Upper.} Spatial profiles of emissions for co-measured H$_2$O (blue), CO (green), and continuum (dashed) in E3, with traces and labels as in Figure~\ref{fig:profiles1}. The slit was oriented along the projected Sun-comet line. \textbf{Lower.} Spatial profiles of co-measured H$_2$O, CO, and continuum in E3 with the slit oriented perpendicular to the projected Sun-comet line.  
\label{fig:profiles2}}
\end{figure*}

\begin{figure*}
\plotone{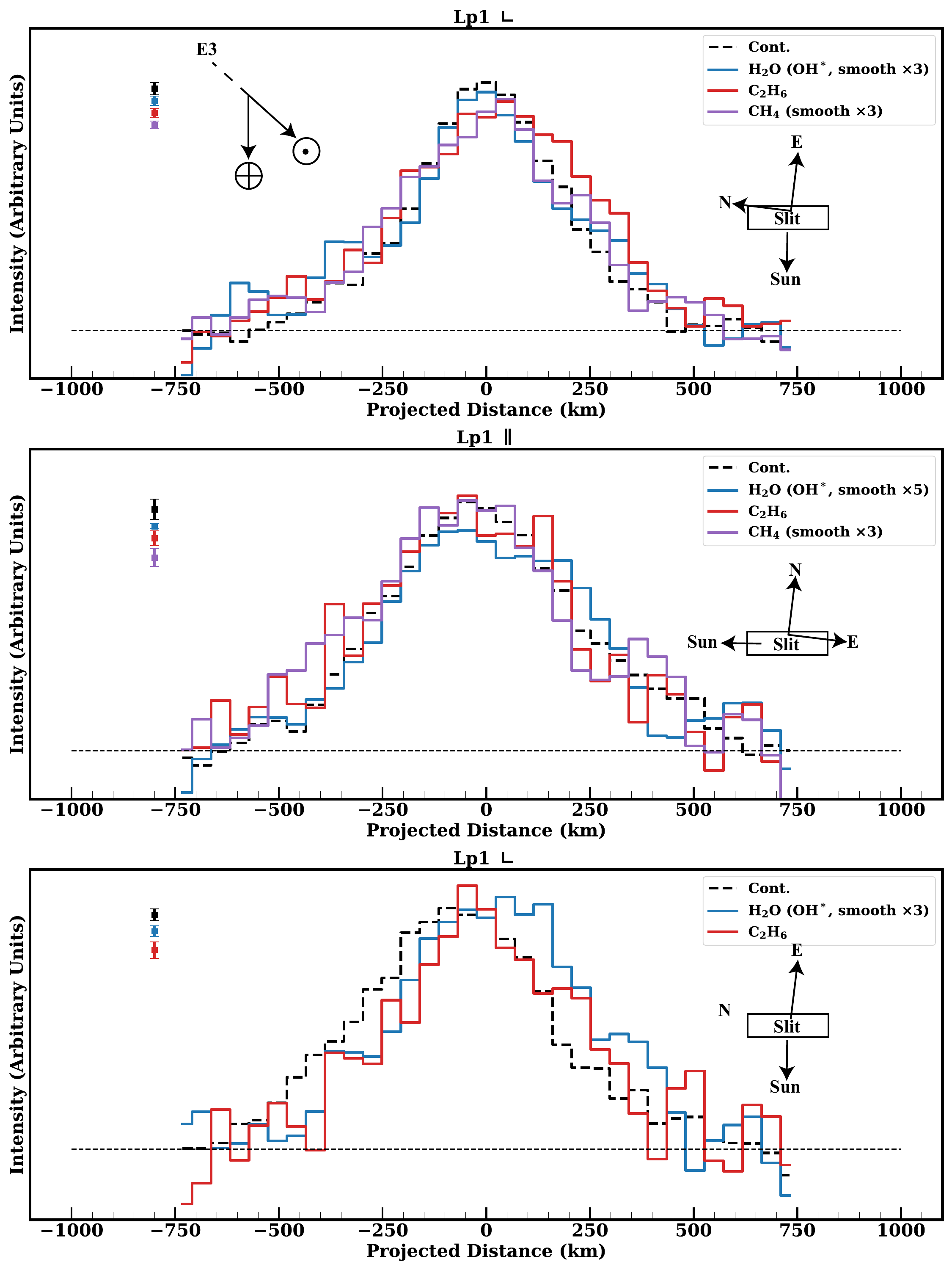}
\caption{\textbf{Upper.} Spatial profiles of emissions for co-measured H$_2$O (OH$^*$, blue), C$_2$H$_6$ (red), CH$_4$ (purple), and continuum (dashed) in E3, with traces and labels as in Figure~\ref{fig:profiles1}. The slit was oriented perpendicular to the Sun-comet line (first integration, UT 07:55 - 08:40). \textbf{Middle.} Spatial profiles of H$_2$O, C$_2$H$_6$, CH$_4$ and continuum in E3 with the slit oriented along the projected Sun-comet line. \textbf{Lower.} Spatial profiles of H$_2$O, C$_2$H$_6$, CH$_4$ and continuum in E3 with the slit oriented perpendicular to the projected Sun-comet line (third integration, UT 09:39 - 10:15).
\label{fig:profiles3}}
\end{figure*}

We were able to extract spatial profiles of emissions for H$_2$O \citep[measured directly or through its proxy, OH*; see][]{Bonev2006}, HCN, CO, C$_2$H$_6$, and CH$_4$ in E3 (Figures~\ref{fig:profiles1}, \ref{fig:profiles2}, \ref{fig:profiles3}). Although OH* has been established as a reliable tracer for the production and spatial distribution of its parent, H$_2$O, it is important to note in the context of this study that spatial profiles of the two molecules may show subtle differences, particularly in the case of H$_2$O sublimation from icy grains. The high brightness and small geocentric distance of E3 enabled us to map the inner coma (within $\sim$750 km to either side of the nucleus) in high detail, with the $0\farcs167$ iSHELL spatial pixel scale subtending a projected distance of 45 km. The moderate phase angle ($\phi \sim$ 48$\degr$) during our observations must be considered when interpreting spatial profiles. 

\begin{deluxetable*}{cccccc}
\tablenum{3}
\tablecaption{Emission Spatial Profile Characteristics in C/2022 E3\label{tab:profiles}}
\tablewidth{0pt}
\tablehead{
\colhead{Setting/Orientation} & \colhead{Molecule} & \colhead{GF$^{(a)}$} & \multicolumn{2}{c}{Differences (Volatiles - Dust)} & \colhead{Asymmetry$^{(d)}$} \\ \cline{4-5}
\colhead{} & \colhead{} & \colhead{} & \colhead{HWHM$^{(b)}$ (pixels)} & \colhead{Peak Position$^{(c)}$ (pixels)} & \colhead{}
}
\startdata
L-Custom & H$_2$O & $2.21\pm0.06$ & $0.50\pm0.61$ & $-0.22\pm0.94$ & $0.76\pm0.03$ \\
PA = 88$\degr$ & HCN & $2.05\pm0.22$ & $0.00\pm0.62$ & $2.61\pm0.99$ & $5.69\pm0.75$ \\
\hline
L-Custom & H$_2$O & $2.38\pm0.02$ & $0.50\pm0.56$ & $0.08\pm0.78$ & $1.00\pm0.01$ \\
PA = $178\degr$ & HCN & $1.73\pm0.21$ & $-1.50\pm0.67$ & $-2.48\pm0.87$ & $0.20\pm0.03$ \\
\hline
M2 & H$_2$O & $2.03\pm0.15$ & $0.50\pm0.64$ & $-0.99\pm0.82$ & $0.63\pm0.04$ \\
PA = 178$\degr$ & CO & $2.06\pm0.15$ & $1.00\pm0.83$ & $-0.10\pm1.09$ & $1.29\pm0.03$ \\
\hline
M2 & H$_2$O & $2.37\pm0.11$ & $1.00\pm0.86$ & $0.47\pm1.26$ & $1.09\pm0.05$ \\
PA = 88$\degr$ & CO & $2.42\pm0.14$ & $1.50\pm0.76$ & $-0.78\pm1.06$ & $0.63\pm0.02$ \\
\hline
Lp1 & H$_2$O (OH$^*$) & $2.01\pm0.12$ & $1.00\pm1.02$ & $0.12\pm1.64$ & $0.82\pm0.03$ \\
PA = 178$\degr$ & CH$_4$ & $2.27\pm0.12$ & $1.00\pm0.85$ & $-0.02\pm1.51$ & $1.41\pm0.04$ \\
 & C$_2$H$_6$ & $2.54\pm0.05$ & $1.50\pm0.73$ & $0.83\pm1.27$ & $0.93\pm0.02$ \\
 \hline
 Lp1 & H$_2$O (OH$^*$) & $2.77\pm0.15$ & $0.50\pm1.08$ & $0.60\pm1.71$ & $0.50\pm0.01$ \\
 PA = 88$\degr$ & CH$_4$ & $2.50\pm0.28$ & $0.50\pm1.34$ & $-0.59\pm2.24$ & $0.83\pm0.03$ \\
  & C$_2$H$_6$ & $2.56\pm0.08$ & $0.50\pm1.33$ & $-0.10\pm1.98$ & $0.78\pm0.03$ \\
\hline
Lp1 & H$_2$O (OH$^*$) & $2.32\pm0.15$ & $0.00\pm1.60$ & $2.24\pm3.89$ & $1.09\pm0.03$ \\
PA = 178$\degr$ & C$_2$H$_6$ & $2.32\pm0.08$ & $-0.50\pm1.81$ & $1.37\pm4.06$ & $0.46\pm0.03$
\enddata
\tablecomments{\sups{a}Growth factor. This represents the ratio of global \textit{Q} to \textit{Q}\subs{NC} based on the central $0\farcs75\times0\farcs83$ aperture having peak emission flux. \sups{b}Difference in spatial profile full width at half maximum between volatile emission and co-measured in dust (1 pixel represents a projected distance at the comet of 45 km). Positive values correspond to a broader volatile profile compared to the dust, whereas negative values indicate a narrower profile. \sups{c}Difference between the spatial profile peak position between volatile emission and co-measured dust. For measurements taken along the projected Sun-comet vector (PA = 88$\degr$), positive values correspond to a sunward (westward) volatile shift, whereas for measurements taken perpendicular (PA = 178$\degr$), positive values correspond to a northward volatile shift. \sups{d}Symmetry of the profile as measured by the ratio of flux along each side of the slit, excluding the central 5 pixels around the peak flux position.
}
\end{deluxetable*}

Spatial profiles of emission provide insights into how molecules are produced: those peaking strongly near the nucleus and falling off with $\sim\rho^{-1}$ dependence are associated with direct nucleus release, whereas molecules with a very broad, flat spatial distribution are consistent with daughter products or extended source molecules \citep{DelloRusso2016b}. For molecules produced via direct nucleus sublimation, spatial profiles reveal whether they originate from common or distinct outgassing sources on the nucleus, and in turn, whether they were associated or segregated as ices in the nucleus. In particular, molecules with distinct outgassing sources on the nucleus should project differently onto the slit as its orientation (position angle) is changed.

\cite{DelloRusso2022} characterized the spatial profiles in comet C/2014 Q2 (Lovejoy) using four quantities: (1) the multiplicative growth factor, (2) the difference in half width at half maximum (HWHM) of a Gaussian fit to the volatile emission profile as compared to that of co-measured dust, (3) the difference in peak pixel of the volatile emission (determined from the peak of the Gaussian fit) compared to that of co-measured dust, and (4) the symmetry of the volatile emission profile derived from the ratio of summed flux along each side of the slit (for E3, this was from $\sim150$ km to $\sim$750 km projected distance). 

We followed a similar procedure for E3; however, given the clear asymmetry of some of the profiles (HCN in particular), we instead performed least squares fitting of a skewed Gaussian to the profiles using the \texttt{SkewedGaussianModel} of the \texttt{lmfit} application. This skewed Gaussian allowed for a considerably improved fit to the profiles compared to a classical Gaussian. For measurements taken with a slit position angle oriented parallel to the projected Sun-comet line (PA = 88$\degr$), quantities (3) and (4) are with respect to the sunward/anti-sunward directions, whereas for a perpendicular slit position angle they are with respect to the roughly north/south directions. Our results are quantified in Table~\ref{tab:profiles}.

During the L-custom setting measurements along the projected Sun-comet line, H$_2$O emission was relatively symmetric, while being only slightly broader than and peaking co-spatially with the co-measured dust. In contrast, HCN was highly asymmetric, peaking in the anti-sunward direction while showing a sunward enhancement. When measured perpendicular to the Sun-comet line, the characteristics of the H$_2$O profile were largely the same, although it was more symmetric than the sunward/anti-sunward projection. Likewise, HCN continued to show considerable asymmetry, yet the sense of the asymmetry and the direction of the peak pixel offset changed. Given these significant changes in the HCN spatial profile between slit orientations (compared to the consistency of the H$_2$O), it is likely that the E3 displayed heterogeneous outgassing, with each molecule originating from distinct nucleus sources.

During observations with the M2 setting, H$_2$O and CO emission were both relatively symmetric and co-spatial with the continuum; however, H$_2$O peaked in the sunward direction, whereas CO peaked in the projected anti-sunward direction. During the Lp1 setting, H$_2$O emission (measured through its proxy, OH$^*$ prompt emission) was co-spatial with the dust during the first measurement perpendicular to the Sun-comet line and also the measurement parallel to it. However, it drifted considerably from the continuum during the second perpendicular measurement. These differences are notable in the context of spatial profiles of H$_2$O and OH* measured in C/2014 Q2 Lovejoy, where H$_2$O peaked co-spatially with and was only slightly broader than the dust, whereas the OH* profiles showed an anti-sunward offset and broader profiles compared to the dust \citep{DelloRusso2022}. Similarly, C$_2$H$_6$ showed changes, being broader than H$_2$O during the first perpendicular measurement yet narrower than it in the second, and its asymmetry also changed. In contrast, CH$_4$ was co-spatial with the dust and relatively symmetric in all instances. Overall, these spatial profiles suggest common outgassing sources for H$_2$O, CO, and CH$_4$, whereas HCN and potentially C$_2$H$_6$ may have had distinct sources.

%% file: results-temperature.tex
\subsection{Spatial Profiles of H$_2$O Rotational Temperature}\label{subsec:tprofiles}
\begin{figure*}
\plotone{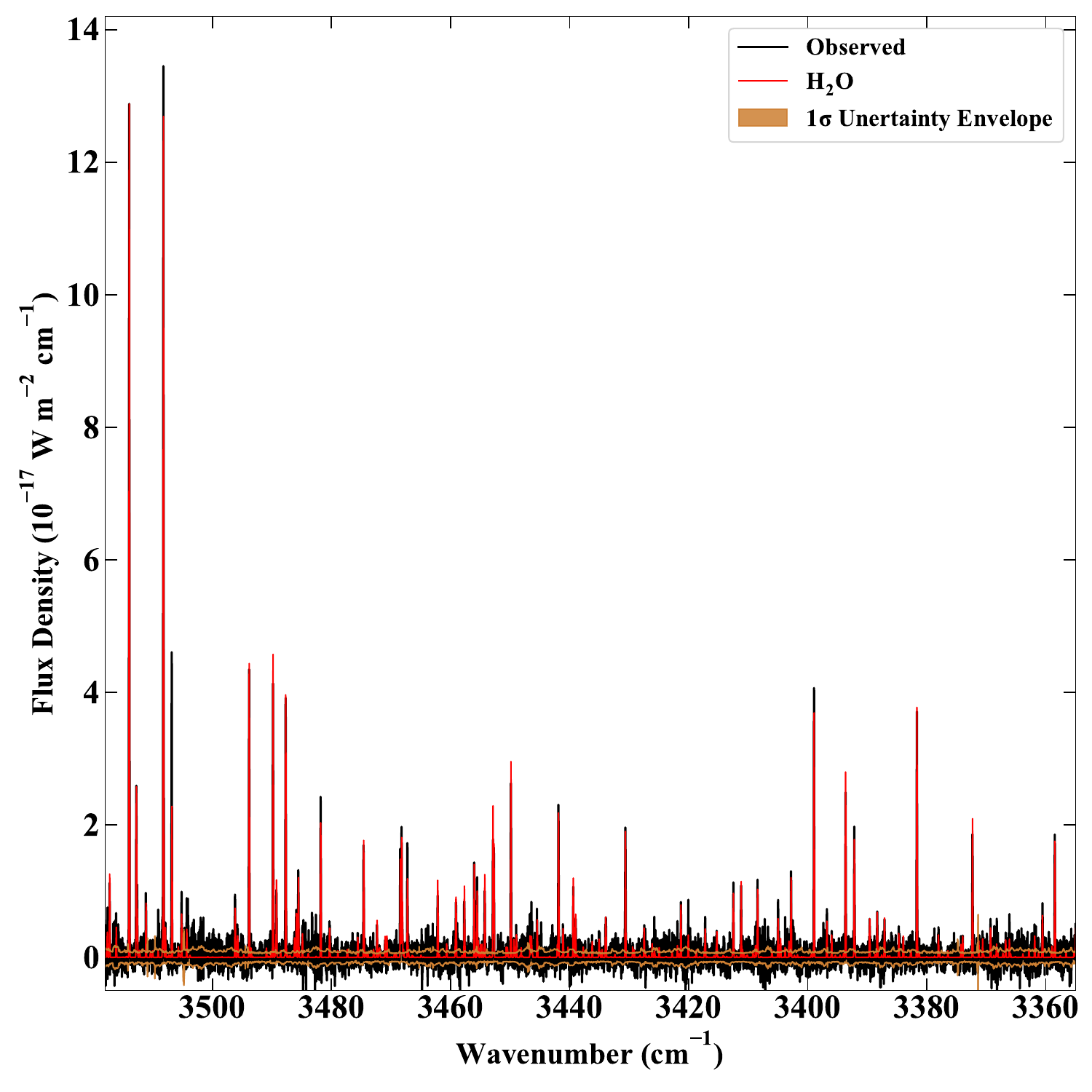}
\caption{ Detections of H$_2$O in E3 covering iSHELL echelle orders 174--182. The best-fit fluorescence model is overlaid in red, and the $1\sigma$ uncertainty envelope is shaded in bronze. 
\label{fig:all-h2o}}
\end{figure*}

\begin{figure*}
\plotone{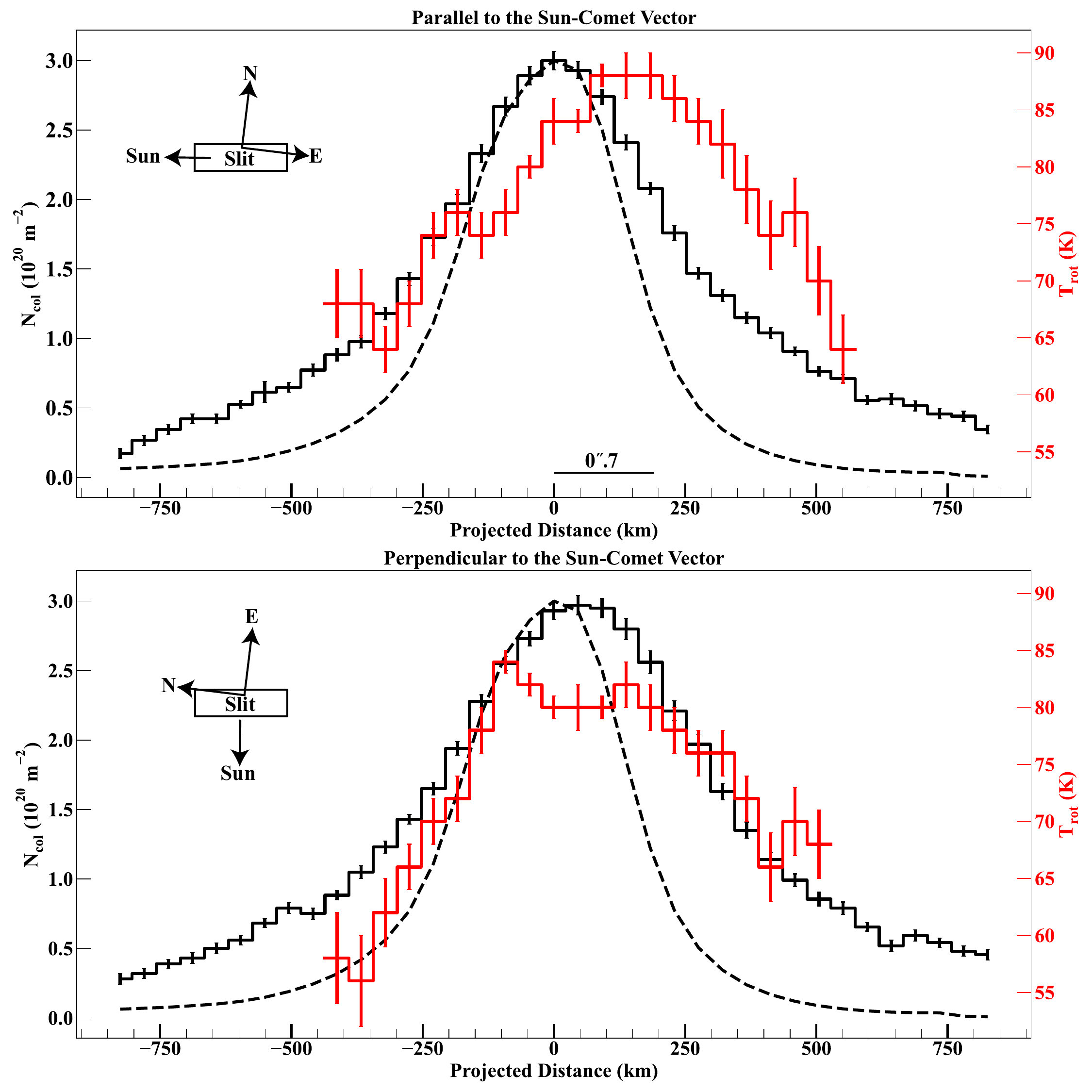}
\caption{\textbf{Left.} Spatial profiles of column density (black) and rotational temperature (red) for H$_2$O in E3 with the slit positioned along the projected Sun-comet line. The stellar PSF (dashed) is also shown, along with the orientation of the slit. The black bar indicates the $0\farcs7$ seeing. \textbf{Right.} Spatial profiles of H$_2$O column density and rotational temperature in E3 with the slit oriented perpendicular to the projected Sun-comet line.
\label{fig:tprofiles}}
\end{figure*}

Trends in rotational temperature along the slit can be tested for molecules with sufficiently high signal-to-noise at off-nucleus positions. For E3, this was satisfied for H$_2$O. We extracted spectra in sliding 1-pixel ($0\farcs167$) extracts along the slit and retrieved molecular column density and rotational temperature at each position. Figure~\ref{fig:all-h2o} shows all H$_2$O transitions detected in a nucleus-centered extract of a single 40 minute on-source integration, and Figure~\ref{fig:tprofiles} shows the extracted column density and rotational temperature profiles.

The rotational temperature profiles for each slit orientation are distinct and asymmetric about the nucleus position. The temperature profile measured along the projected Sun-comet line is $84\pm2$ K at the nucleus position and rises to $88\pm2$ K at $\sim$135 km off-nucleus in the anti-sunward direction. Although the temperature in the projected sunward side falls to $\sim$65 K within 320 km, the temperature in the anti-sunward side does not reach the same value until nearly 550 km from the nucleus. 

The temperature profile measured perpendicular to the projected Sun-comet line is also remarkable, rising from $80\pm1$ K at the nucleus position to $84\pm1$ K at 90 km north of the nucleus before beginning to decrease. There is also a difference in heating between the northern and southern hemispheres, with the north side cooling to $66\pm2$ K within 270 km of the nucleus, whereas the southern hemisphere doesn't reach the same value until 410 km from the nucleus. Collectively, these results indicate significant differences in the balance of heating and cooling mechanisms in the day and night sides of the coma.

%% file: Discussion.tex
\section{Discussion}\label{sec:discussion}
Our measurements of E3 detail its complex coma, revealing distinct sources of outgassing for different species, asymmetric coma temperature profiles, and molecular abundances for multiple species. We address each of these topics, compare our measurements against those in other studies of E3, and place our results into context with the comet population.

\input{discuss-e3}

%% file: discuss-e3.tex
\subsection{Molecular Abundances in C/2022 E3}\label{subsec:disc-abund}
Our measurements in C/2022 E3 provide a record of its coma composition over $\sim$7 hours near its closest approach to Earth. On the one hand, overall coma activity varied significantly, with $Q$(H$_2$O) falling nearly 30\% from its value during our first Lcustom block (UT 03:07--03:52) to its value during the first M2 integration (UT 05:38--05:50), then rising nearly the same amount by the second Lp1 block (UT 08:47--09:32; Table~\ref{tab:comp}). On the other hand, the relative molecular abundances for HCN, CO, and CH$_4$ agreed within 2$\sigma$ uncertainty for all integrations of a given instrumental setting. In this context, it is also interesting to emphasize that despite clear evidence for heterogenous outgassing of HCN and H$_2$O, their relative molecular abundances differed by $<2\sigma$ uncertainty between the first and second L-Custom integrations. This scenario of flat molecular abundances despite complex outgassing patterns and changing slit orientations is consistent with results seen in EPOXI target 103P/Hartley 2 \citep{DelloRusso2011,Mumma2011b,Kawakita2013}.

In contrast, CH$_3$OH and C$_2$H$_6$ abundances decreased by 30\% and 45\%, respectively, between the first and second Lp1 observations before moderately increasing during the third. The trend in CH$_3$OH/H$_2$O and C$_2$H$_6$/H$_2$O can be explained in terms of a continuous decrease in $Q$(CH$_3$OH) and $Q$(C$_2$H$_6$) across the entire Lp1 sequence while $Q$(H$_2$O) cycled up and down. On the other hand, despite significant variation in the CH$_4$/C$_2$H$_6$ abundances, CH$_3$OH/C$_2$H$_6$ remained constant within uncertainty. It is therefore appropriate to recall that H$_2$O was not measured directly in the Lp1 setting (rather through its proxy, OH$^*$), and that overall, relative abundances of molecules measured simultaneously within a setting (i.e., HCN, CO, C$_2$H$_2$, and H$_2$O; CH$_3$OH, and CH$_4$) showed the most consistency across time and are likely the most robust. 

Despite the uncertainties, it is possible that the variation in $Q$(H$_2$O) was tied to rotational effects, as our lowest values coincide with measurements taken halfway through its $\sim$8 hour rotational period \citep{Manzini2023,Knight2023}, with different active sites rotating into and out of the dayside hemisphere and dominating activity. \cite{Foster2026} found evidence for such variations in $Q$(H$_2$O) through spatial-spectral mapping of H$_2$O with JWST roughly one month after this study. Observations at radio wavelengths produced similar results, with \cite{Biver2024a} reporting variations in H$_2$O and CH$_3$OH production rates with an 8--9 hour period, and \cite{Li2025} finding that HCN was temporally variable as well. 

We can also compare our molecular abundances against those found for E3 in other studies at infrared and radio wavelengths. Our abundances for HCN and CO are in formal agreement with those found by \cite{Biver2024a}, but lower than those reported in \cite{Li2025}, and both molecules are considerably depleted with respect to average values in measured Oort cloud comets \citep{DelloRusso2016}. It is interesting to note that HCN abundances in a given comet often disagree between radio and infrared measurements \citep[by up to a factor of 6;][]{Biver2024b}, and E3 is an uncommon example of good agreement. On the other hand, our CH$_3$OH abundance is lower by a factor of $\sim$2--3 than that reported by \cite{Biver2024a} and a factor $\sim1.7$ than \cite{Foster2026}, yet all values are depleted compared to average values in Oort cloud comets \citep{DelloRusso2016}. Conversely, our OCS abundance is twice as high as \cite{Biver2024a} yet in agreement with \cite{Foster2026}, and consistent with its mean value among measured comets \citep{Saki2020}.

As radio observations cannot sample the symmetric hydrocarbons, which lack a rotational dipole moment, our study provides their sole report in E3 to date. Our C$_2$H$_6$, CH$_4$, and C$_2$H$_2$ abundances are all depleted compared to their mean abundances among measured Oort cloud comets \citep{DelloRusso2016}. 

\subsection{Coma Thermal Physics in C/2022 E3}\label{subsec:disc-temp}
Our H$_2$O rotational temperature profiles (Figure~\ref{fig:tprofiles}) measured at distinct position angles in the coma provide a window into the coma physics of E3. The rotational temperature profiles did not track the column density profiles in either slit orientation, and in each case distinct hemispheric asymmetries were observed, with the rotational temperature peaking off-nucleus and cooling occurring significantly slower in one hemisphere. In particular, the profile measured parallel to the Sun-comet line was cooler and decreased more quickly in the projected sunward direction, while peaking off-nucleus and cooling more slowly in the anti-sunward side. The disparity in peak position between the H$_2$O rotational temperature and column density profiles measured parallel to the Sun-comet line is consistent with the effects of icy grains being dragged by radiation pressure into the anti-sunward hemisphere before releasing their H$_2$O at an elevated \trot{} (namely the vacuum sublimation temperature of H$_2$O). This interpretation is also consistent with the double-peaked temperature profile measured perpendicular to the Sun-comet line, where the anti-sunward, off-nucleus rise in rotational temperature is projected symmetrically along the line of sight in the north and south directions.

These results are similar to those found in ground-based observations of EPOXI target 103P/Hartley 2 \citep{Bonev2013}, with the H$_2$O rotational temperature profile along the Sun-comet line decoupled from the column density profile and an increase in temperature in the anti-sunward direction. Measurements by the EPOXI spacecraft demonstrated a complex outgassing geometry originating from the nucleus of 103P/Hartley 2, with icy H$_2$O-coated grains dragged into the coma from the waist region before subliming and adding to the coma water content. This synergy between ground- and space-based results provided strong evidence for the diagnostic signature of icy grain sublimation in high-resolution ground-based spectroscopy as reported here for E3.

Our results are also consistent with those found in comet 46P/Wirtanen at radio wavelengths using single dish spectroscopy \citep{Biver2021} and interferometric imaging with the ALMA array \citep{Cordiner2023}, which were explained by the combined effects of more efficient adiabatic cooling in the dayward side and the heating effects of sublimation from icy grains in the anti-sunward side. It is important to note that both 103P/Harley 2 and 46P/Wirtanen were hyperactive comets, with their overabundant H$_2$O production relative to their small nucleus size attributed to icy grain sublimation in the coma.

To further test the likelihood of icy grain sublimation in E3, we worked to determine whether it was a hyperactive comet like 46P/Wirtanen or 103P/Hartley 2. A hyperactive comet is defined as having a nucleus active fraction greater than 100\%. \cite{Liu2024} found a lower limit to E3's nucleus size of 0.87 km and an upper limit of 2.79 km. We calculated the H$_2$O active fractional area for E3 using the sublimation model of \cite{Cowan1979} for a visual abledo of 0.05 and an infrared emissivity of 0.95. We used the Small-Bodies-Node/ice-sublimation code \citep{VanSelous2021} to calculate the average H$_2$O sublimation rate per surface unit, $Z$, at \rh{} = 1.05 au. We used the absolute mean $Q$(H$_2$O) during our observations of $(7.5\pm0.9)\times10^{28}$ \ps{}. The active area is calculated by dividing $Q$(H$_2$O) by $Z$, and the nucleus active fractional area is found by dividing the active area by the nucleus surface area. We find $Z=3.25\times10^{17}$ mol cm$^{-2}$ s$^{-1}$. For a nucleus radius of 2.79 km and our $Q$(H$_2$O), this corresponds to an active fraction of 28\%, whereas for a 0.87 km nucleus radius the active fraction is 279\%. Indeed, if E3's nucleus radius were smaller than 1.35 km, it would be a hyperactive comet during our observations. Combined with our rotational temperature and column density profiles, this is another line of evidence that icy grain sublimation likely played a significant role in E3's coma near its closest approach to Earth.

However, the thermal physics of E3's coma did not necessarily remain constant with time. \cite{Foster2026} also found generally higher rotational temperatures in the projected anti-sunward hemisphere of E3 roughly one month (2023 March 1 - 4) after the observations reported here. This sunward/anti-sunward dichotomy was seen for both H$_2$O (measured with JWST NIRSpec IFU observations) as well as CH$_3$OH (measured with ALMA), with both instruments sampling gas within $\sim1000$ km projected distance from the nucleus. The higher anti-sunward temperature was explained as a consequence of more efficient adiabatic cooling in the sunward hemisphere. However, their observations showed a continuous decrease of the rotational temperature with distance in both the projected sunward and anti-sunward hemispheres, without any evidence for an off-nucleus increase in temperature. Thus, it is possible that icy grain sublimation was not a constant effect in E3's coma.

Aside from icy grain sublimation, off-nucleus increases in temperature may also be explained by the effects of electron collisional heating \citep{Biver1999,Cordiner2025}. However, \cite{Foster2026} demonstrated through detailed radiative transfer modeling that such heating should have taken place near the electron contact surface at roughly 750 km projected distance from the nucleus for E3. This is considerably more distant than the off-nucleus temperature increases in this work, which occur at $\sim100-200$ km projected distance, leaving icy grain sublimation as the preferred explanation for our measurements.

%% file: Conclusions.tex
\section{Conclusion}\label{sec:conclusion}
Our observations of comet C/2022 E3 near its closest approach to Earth leveraged the high resolving power and wide spectral grasp of the NASA-IRTF to reveal its coma composition and thermal physics with high precision. Our results found a comet that showed variation in its overall activity level that was likely tied to the rotation of its nucleus, yet molecular abundances remained largely consistent on 2--3 hour timescales. Compared to mean values among measured Oort cloud comets, its molecular composition measured at IR wavelengths was overall depleted, consistent with studies at radio wavelengths. We found that H$_2$O, CO, and CH$_4$ likely shared common outgassing sources based on spatial profiles of their emission, whereas HCN and potentially C$_2$H$_6$ may have had distinct sources. Analyzing rotational temperature profiles of H$_2$O, we found evidence for a complex temperature structure, with the rotational temperature profiles decoupled from those of the column density. These may indicate the presence of H$_2$O sublimation from icy grains in the coma, and our calculations of the nucleus active fraction suggests that E3 may have been a hyperactive comet. Together these results highlight the premier capabilities of the NASA-IRTF for disentangling the complex coma chemistry and physics seen in objects such as C/2022 E3 ZTF.